# Photon echo quantum RAM integration in quantum computer

Sergey A. Moiseev [1,2 *], & Sergey N. Andrianov [1,2]

[1] *Kazan Physical-Technical Institute of the Russian Academy of Sciences*
*10/7 Sibirsky Trakt, Kazan, 420029, Russia*

[2] *Institute for Informatics of Tatarstan Academy of Sciences,*
*20 Mushtary, Kazan, 420012, Russia*

E-mail: * *samoi@yandex.ru*

We have analyzed an efficient integration of the multi-qubit echo quantum memory into the quantum computer scheme on the atomic resonant ensembles in quantum electrodynamics cavity. Here, one atomic ensemble with controllable inhomogeneous broadening is used for the quantum memory node and other atomic ensembles characterized by the homogeneous broadening of the resonant line are used as processing nodes. We have found optimal conditions for efficient integration of multi-qubit quantum memory modified for this analyzed physical scheme and we have determined a specified shape of the self temporal modes providing a perfect reversible transfer of the photon qubits between the quantum memory node and arbitrary processing nodes. The obtained results open the way for realization of full-scale solid state quantum computing based on using the efficient multi-qubit quantum memory.

**I. INTRODUCTION**

Construction of multi-qubit quantum computer (QC) is a genuine challenge for modern quantum physics and quantum engineering that imposes various critical requirements for organization of controlled highly delicate quantum dynamics of qubits [1,2]. Quantum computation exploits a number of programming single- and two- qubit gates for the qubits incorporated in the QC where the quantum processing demands for an easy address to arbitrary qubits and convenient coupling of each pair of the qubits. These and other physical demands to the QC have to be fulfilled with almost 100 % fidelity and efficiency for all the quantum gates [3].

QC architecture must be scalable. Thus, solid state nanotechnologies seem to be promising for the construction of quantum computer. The first two-qubit solid state prototype of QC was constructed in 2009 on the superconducting elements [4]. Later, three superconducting quantum bits where entangled [5,6] and the prototype of superconducting QC with quantum random access memory (QuRAM) were demonstrated [7]. QuRAM must have much longer decoherence times than that of processor qubits. Therefore the hybrid approach where supercoducting qubits are used for processing and spin qubits are used for storage [8,9] is perspective.

All the requirements on high figure of merit have to be satisfied not only for processing nodes of QC, but also, for multi-qubit *quantum memory* (QM) integrated in the QC circuit. Two of four purified De'Vinzenco criteria on QC concern QM [10]. QM must serve as input and output terminal of QC allowing, also, temporal storage of qubits and entanglement. Once efficient multimode QM is created, the scalability will be achieved for QC with limited resources of processor unit. Moreover, availability of QM in QC will allow parallel quantum computing.

Photon echo based QM of 69% efficiency has been demonstrated recently by using a solid state medium that had beaten 50% threshold of quantum efficiency [11]. However the QC requires the QM of almost 100% efficiency for storage of the multi-qubit states that will be a huge problem for the nearest future investigations. First experiments have been done recently towards realization of multi-qubit QM in the QC [12]. Another relating problem is an integration of the QM in the QC for effective coupling with processing gates and quantum communication with external environment of the QC.

Recently we have proposed an efficient multi-qubit photon echo based QM (PEQM) in single mode QED resonator (cavity PEQM) [13,14,15]. Also, it was proposed in paper [16] without spectral matching condition. The proposed PEQM provides desirable (almost 100 %) quantum efficiency in a perfect retrieval of the stored quantum states for multi-qubit fields using *moderate physical parameters* of atomic system in the optimal QED cavity and controlled inhomogeneously broadened resonant transition [17-21]. A considerable progress in the realization of the efficient coupling with atomic ensembles in resonators have been recently demonstrated in [22-24].

Here, we elaborate principles for the integration of the cavity PEQM in the QC architecture providing a perfect storage and reversible transfer of photon qubits stored in the QM node to processing nodes. Finally, we summarize the obtained results and discuss other topic problems of the QM integration in the quantum processing.

**II. INTEGRATED MULTI-MODE QUANTUM MEMORY**

For analysis of the QM, we use the quantum electrodynamics of the atomic ensembles in the single mode QED cavity with following cavity mode formalism [25]. We use two-level atomic ensembles for modeling the QM node and processing nodes in their resonant interaction with cavity mode field. We take the generalized Tavis-Cumming Hamiltonian [26,18] for the atomic systems including the QM node with the processing gates and field modes, where

$$H_0 = \hbar\omega_0 \left\{ \sum_{m=0}^{M} \sum_{j_m=1}^{N_m} S_z^{j_m} + a_\sigma^+ a_\sigma + \sum_{n=1}^{2} \int b_n^+(\omega) b_n(\omega) d\omega \right\}, \quad (1)$$

is a basic Hamiltonian containing the main energies of atoms ($S_z^{j_m}$ is a z-projection of the spin operator in m-th node, m=0 corresponds to QM node, M is a number of processing nodes, Nm is number of atoms in m-th node), energy of cavity σ-field mode (where $a_\sigma^+(\omega)$ and $a_\sigma(\omega)$ are creation and annihilation operators), energy of waveguide field (n=1) and energy of free space field (n=2) ($b_n^+$ and are $b_n$ creation and annihilation operators of the waveguide modes), generalized by taking into account inhomogeneous broadening of atomic frequencies and continuous spectral distribution of the field modes

$$H_1 = \\
= \hbar \sum_{m=0}^{M} \sum_{j_m}^{N_m} \left( \Delta_{j_m}(t) + \delta\Delta_{j_m}(t) \right) S_z^{j_m} + \hbar \sum_{n=1}^{2} \int (\omega - \omega_0) b_n^+(\omega) b_n(\omega) d\omega + \\
+ i\hbar \sum_{m=0}^{M} \sum_{j_m}^{N_m} \left( g_{j_m} S_-^{j_m} a^+ - g_{j_m}^* S_+^{j_m} a \right) + i\hbar \sum_{n=1}^{2} \int \kappa_n(\omega) \left[ b_n(\omega) a^+ - b_n^+(\omega) a \right] \quad (2)$$

is a perturbation Hamiltonian.

first term in (2) comprises perturbation energies of atoms where $\Delta_{j_m}(t)$ is a controlled frequency detuning of j-th atom; $\Delta_{j_0}(t<\tau) = \Delta_{j_0}$ and $\Delta_{j_0}(t>\tau) = -\Delta_{j_0}$ for QM node in the analyzed storage protocol; $\delta\Delta_j(t)$ is its fluctuating frequency detuning determined by local stochastic fields, $g_{j_m}$ is a coupling photon-atom constant. Ensemble distributions over the detunings $\Delta_{j_0}(t)$ and $\delta\Delta_{j_m}(t)$ determine the inhomogeneous and homogeneous broadenings of the resonant atomic lines in the atomic nodes. We assume that all atoms in the processing nodes (m=1,…,M) have equal frequency detunings ($\Delta_{j_m}(t) = \Delta_m(t)$). In the following, we use Lorentzian shape for inhomogeneous broadening of the QM node and typical anzatz for ensemble average over the fluctuating detunings) $\delta\Delta_j(t)$:

$$\sum_{j_m=1}^{N_m} \left| g_{j_m} \right|^2 \exp\left\{ -i\Delta_{j_m}(t-t') \right\} \Phi_{j_m}(t,t') \equiv N_m \left| \bar{g}_m \right|^2 \exp\left\{ -(\Delta_{in} + \gamma_{21}) |t-t'| \right\}, \quad (3)$$

where $\Phi_{j_m}(t,t') = \exp\{i\varphi_{j_m}(t,t')\}$, $\varphi_{j_m}(t,t') = \int_{t'}^{t} d t'' \delta\Delta_{j_m}(t'')$, $\gamma_{21}$ is a spectral width of the homogenous Lorentzian line, $|\bar{g}_m|^2$ is a quantity averaged over the atoms in m-th node. Second term in (2) contains frequency detunings of the field n-th modes. Third terms is the interaction energy of atoms with cavity mode ($S_+^{j_m}$ and $S_-^{j_m}$ are the transition spin operators). Fourth term is an interaction energy of the cavity mode with the waveguide and free propagating modes characterized by coupling constants $\kappa_n(\omega')$.

We note that $[H_0, H_1] = 0$ and Hamiltonian $H_0$ characterizes a total number of excitations in the atomic system and in the fields which is conserved during the quantum evolution where $H_0$ gives a contribution only to the evolution of common phase of the wave function. $H_1$ determines a unitary operator causing a coherent evolution of the atomic and field systems with dynamical exchange and entanglement of the excitations between them. In spite of huge complexity of the compound light-atoms system, we show here that their quantum dynamics governed by $H_1$ can be perfectly reversed in time on our demand in a simple robust way. We assume that initially all atoms in m-th node ($j=1,2,...,N_m$) stay on the ground state $|0\rangle_a = |0_1, 0_2, ..., N_m\rangle$ (m=1,2,...M) and we launch a signal multi-mode single photon fields prepared in the initial quantum state $|\psi_{in}(t)\rangle_{ph} = \prod_{k=1}^{N_{ph}} \psi_k^+(t-t_k)|0\rangle$, $\psi_k^+(t-t_k) = \int_{-\infty}^{\infty} d\omega_k f_k(\omega_k) \exp\{-i(t-t_k)\} b_1^+(\omega_k)$, $\int_{-\infty}^{\infty} d\omega_k |f_k(\omega_k)|^2 = 1$ ($N_{ph}$ is a number of modes) to the optical cavity that is a vacuum state of the field. k-th mode arrives in the circuit at time moment $t_k$, time delays between the nearest photons are assumed to be large enough $(t_k - t_{k-1} >> \delta t)$, $\delta t$ is a temporal duration of the photon wave packets. Additional free field modes (n=2) are assumed to be in the vacuum state $b_2(\omega) \cdot |0\rangle = 0$. Thus the total initial state of light and atoms is $\Psi_{in}(t) = |\psi_{in}(t)\rangle_{ph} |0\rangle_a$.

Neglecting a population of excited states of the multi-atomic ensembles and using the input and output field formalism [25,14], we derive the following linearized system of Heisenberg equations for the field operators and for the atomic operators of the QM node (m=0) and processing nodes (m=1,...M) in the rotating frame representation:

$$\frac{d}{dt} b_{l;\omega}(t) = -i(\omega - \omega_0) b_{l;\omega}(t) - \kappa_l(\omega) a(t), \qquad (4)$$

$$\frac{d}{dt} S_-^{j_m}(t) = -g_{j_m}^* a - i[\Delta_{j_m} + \delta\Delta_{j_m}(t)] S_-^{j_m}(t), \qquad (5)$$

$$\frac{d}{dt} a(t) = \sum_{m=0}^{M} \sum_{j_m}^{N_m} g_{j_m} S_-^{j_m}(t) - \frac{1}{2}(\gamma_1 + \gamma_2) a(t) + \{\sqrt{\gamma_1} b_{1;in}(t) + \sqrt{\gamma_2} b_{2;in}(t)\}, \qquad (6)$$

where $\gamma_l = 2\pi\kappa_l^2(\omega_0)$, (l=1,2) and m=1 for the QM node. The input signal field containing n temporally separated photon modes is given by $b_{1,in}(t) = \sum_{k=1}^{N_{ph}} b_{1k}(t-t_k)$, where $b_{1k}(t-t_k) = (1/\sqrt{2\pi}) \int_{-\infty}^{\infty} d\omega \exp[-i(\omega - \omega_0)(t-t_k)] b_{1;\omega}$, $t_k$ is a moment of time when the k-th field mode arrives to the QM node from the external waveguide, $[b_{1;\omega'}, b_{1;\omega}^+] = \delta(\omega' - \omega)$.

By using a Fourier transformation for atomic coherences and field modes ($\{S_-^{j_m}(t), a(t), b_{1,2;in}(t)\} = \int_{-\infty}^{\infty} d\nu e^{-i\nu t} \{S_-^{j_m}(\nu), a(\nu), b_{1,2;in}(\nu)\}$) in Eqs. (4)-(6), we find the excited coherences in the processing nodes

$$S_-^m(\nu) = \frac{i g_m^* a(\nu)}{[\Delta_m - \nu - i\gamma_{21}]}, \tag{7}$$

where we replaced index "$j_m$" by the common index "m" in the frequency detunings and in coupling constants. We use a formal solution of Eq. (5)

$$\{\tfrac{1}{2}(\gamma_1 + \gamma_2) - i[\nu + \frac{\Gamma_{tot}}{2[1 - i\nu/\Delta_{tot}]} + \delta_M(\nu)]\} a(\nu) = \{\sqrt{\gamma_1} b_1(\nu) + \sqrt{\gamma_2} b_2(\nu)\}, \tag{8}$$

where $\delta_M(\nu) = \sum_{m=1}^{M} \Omega_m^2 / [\Delta_m - \nu - i\gamma_{21}]$, $\Gamma_{tot} = 2\Omega_o^2 / \Delta_{tot}$ is a photon absorption rate by ensemble of QM node in unit spectral domain, $\Omega_m^2 = N_m |\bar{g}_m|^2$, $\Delta_{tot} = \Delta_{in} + \gamma_{21}$.

Let us assume large enough spectral detunings in comparison with spectral width of the signal fields $\Delta_m \gg \delta\omega_s$ and with the linewidths of the processing nodes $\Delta_m \gg \gamma_{21}$ that simplifies to $\delta_M(\nu) = \delta_M(0) + \nu M(\nu)$, where $\delta_M(0) = \sum_{m=1}^{M} N_m |g_m|^2 / \Delta_m$ and

$$M(\nu) = \sum_{m=1}^{M} \frac{\Omega_m^2}{\Delta_m (\Delta_m - \nu)} \tag{9}$$

determine a permanent and spectrally ($\nu$) dependent frequency shift of the cavity mode due to the interaction with all atoms in the processing nodes. The shift $\delta_M(0)$ can be controlled and reduced to zero by adiabatic changing of the spectral detunings $\Delta_m(t)$ of each m-th atomic ensemble. So, we can put the total shift $\delta_M(0) = 0$ determining thereby a resonant interaction of the cavity mode (for its central spectral component $\nu = 0$) with ensemble of QM atoms that leads to the following formal solution for the cavity field mode

$$a(\nu) = \frac{2\{\sqrt{\gamma_1} b_{1;in}(\nu) + \sqrt{\gamma_2} b_{2;in}(\nu)\}}{\{\gamma_1 + \gamma_2 + \frac{\Gamma_{tot}}{(1 - i\nu/\Delta_{tot})} - 2i\nu \Pi(\nu)\}}. \tag{10}$$

Solely the factor $\Pi(\nu) = [1 + M(\nu)]$ in Eq. (10) characterizes all difference in the interaction between the field and atoms of QM node caused by a presence of the processing nodes. Therefore, one can perform further all the calculations for the quantum storage of the input signal field similarly to [14]. After algebraic calculations we find the storage quantum efficiency as $Q_{ST} = \bar{P}_{ee;1} / \bar{n}_1$, where $\bar{P}_{ee;1} = \sum_{j=1}^{N_0} \langle S_{+;0}^j S_{-;0}^j \rangle$ is an excited level population of atoms after the interaction with last n-th signal field mode (i.e. for t > tn + δt). Total number of photons in the input signal field is $\bar{n}_1 = \int_{-\infty}^{\infty} d\nu \langle b_1^+(\nu; t \to -\infty) b_1(\nu; t \to -\infty) \rangle = \sum_{k=1}^{N_{ph}} \bar{n}_{1;k}$; where $\bar{n}_{1;k} = \int_{-\infty}^{\infty} d\nu \langle b_1^+(\nu; -\infty) b_1(\nu; -\infty) \rangle_k$ is photons in k-th temporal mode, $\langle ... \rangle_k$ is determined by the quantum averaging over the k-th mode state $\psi_k^+(t - t_k)|0\rangle$. By using (10) and (5), we find the atomic coherences in the QM and processing nodes (9), (11). The storage efficiency of the signal field will be given by $Q_{ST}^M = (1/\bar{n}_1) \sum_{k=1}^{N_{ph}} Q_{ST,k}^M \bar{n}_{1,k}$ with the storage efficiency of k-th mode

$$Q_{ST,k}^M = \int_{-\infty}^{\infty} d\nu Z^M(\nu, \Delta_{in}, \gamma_1, \gamma_2, \Gamma_{tot}) \langle b_1^+(\nu) b_1(\nu) \rangle_k / \bar{n}_{1,k}, \tag{11}$$

where modified spectral function ($\Delta_{in} >> \gamma_{21}$ is assumed)

$$Z^M(\nu,\Delta_{in},\gamma_1,\gamma_2,\Gamma_{tot}) = \frac{\Delta_{in}^2}{(\Delta_{in}^2+\nu^2)} \frac{4\gamma_1\Gamma_{tot}}{|\gamma_1+\gamma_2+\Gamma_{tot}/(1-i\nu/\Delta_{in})-2i\nu\Pi(\nu)|^2}, \quad (12)$$

characterizes spectral properties of the quantum storage (we have taken into account the expectation values $\langle b_2^+(\nu)b_2(\nu)\rangle = \langle S_{+,m}^j(t_0)S_{-,m}^j(t_0)\rangle = 0$, m=1,…,M, for the used initial state.

For relatively narrow spectral width $\delta\omega_f$ of the signal field and weak atomic decoherence rate $\gamma_{21}$ in comparison with inhomogeneously broadened width ($\delta\omega_{f,k} \approx \delta t^{-1} < \Delta_{in}+\gamma_{21}$), we get from Eqs. (16), (17):

$$Q_{ST,k}^M(\delta\omega_s << \Delta_{tot},\gamma_1) = \frac{\gamma_1}{\gamma_1+\gamma_2} \cdot \frac{4\Gamma_{tot}/(\gamma_1+\gamma_2)}{[1+\Gamma_{tot}/(\gamma_1+\gamma_2)]^2}. \quad (13)$$

Quantum efficiency $Q_{ST,k}^M$ reaches unity at $\Gamma_{tot}/\gamma_1 = 1$ and $\gamma_2/\gamma_1 << 1$ similarly to the matching condition of QM in resonator without processing nodes [14], that shows a possibility of perfect storage for multi-mode signal field in the presence of M processing nodes (where number of the temporal signal modes is limited by $M_{max} \sim \Delta_{in}\gamma_{21}^{-1}$). Note that $\gamma_1 = \Gamma_{tot}$ is a condition of *perfect optimal matching* between the waveguide modes and the atomic system in the single mode cavity. In this case the signal field containing many temporal modes enters into the circuit and transfers to the QM atomic system in one step without any escape from the cavity similar to simple absorption in a resonator [27,28]. The single step storage of the multi-mode field is possible for inhomogeneously broadened atomic (electron spin) transition providing a perfect absorption for arbitrary temporal profile of the light fields with finite spectral width.

Retrieval of the stored light fields is realized here via using a CRIB procedure [14,17-21] by exploiting symmetry properties of the light atoms dynamics for the absorption stage and echo emission stage. After the perfect storage in accordance with above coupling matching condition, we change a sign of the detunings $\Delta_j \rightarrow -\Delta_j$ at time moment t=$\tau$ by changing a magnetic field polarity similar to recent experiments [11,18,29]. For the case of negligibly weak interaction with the free modes and slow atomic decoherence, i.e. assuming $\gamma_{21} \approx 0$, $\gamma_2/\gamma_1 << 1$, we find that the initial quantum state of the multi-mode signal field will be reproduced at $t = 2\tau$ due to complete unitary reversibility in the irradiated echo signal getting the field spectrum inverted relatively to the central frequency $\omega_0$ in comparison with the original one. By taking into account the atomic decoherence and coupling with free modes similar to the absorption storage, we find the echo field:

$$a_{echo} = -\frac{\exp\{-2\gamma_{21}(t-\tau)\}}{\sqrt{2\pi}\gamma_1}\sum_{k=1}^M \int_{-\infty}^\infty d\nu Z^M(\nu,\Delta_{in},\gamma_1,\gamma_2,\Gamma_{tot})b_{1,k}(\nu)\exp\{i\nu(t+\tau_k-2\tau)\}, \quad (14)$$

where we have assumed weaker decoherence in comparison with the temporal duration of the input light pulses ($\delta t\gamma_{21} << 1$). We find the total number of photons irradiated in the echo signal:

$$n_{echo} = \int d\omega b_{echo}^+(\omega,t>>2\tau)b_{echo}(\omega,t>>2\tau) = \sum_{k=1}^M n_{echo,k}, \quad (15)$$

where

$$\langle n_{echo,k}\rangle = \exp\{-4\gamma_{21}(\tau-\tau_k)\}\int_{-\infty}^\infty d\nu[Z^M(\nu,\Delta_{in},\gamma_1,\gamma_2,\Gamma_{tot})]^2 \langle n_{1,k}(\nu)\rangle. \quad (16)$$

So, the quantum memory efficiency will be

$$E_{QM} = (1/\bar{n}_1)\sum_{k=1}^{N_{ph}} \bar{n}_{1,k} E_{QM,k}, \quad (17)$$

where the total retrieval efficiency of the k-th temporal mode in the echo field pulse is

$$E_{QM,k} = \exp\{-4\gamma_{21}(\tau-\tau_k)\}\int_{-\infty}^{\infty} d\nu \left[Z^M(\nu,\Delta_{in},\gamma_1,\gamma_2,\Gamma_{tot})\right]^2 \langle n_{1,k}(\nu)\rangle/\bar{n}_{1,k}, \qquad (18)$$

where the factor $\exp\{-4\gamma_{21}(\tau-t_k)\}$ describes a destructive influence of the atomic decoherence on the QM efficiency during the storage time $2(\tau-t_k))$ of the k-th mode while $[Z^M(...)]^2$- function describes a spectral properties in the quantum efficiency. By taking into account the first optimal matching condition $\Gamma_{tot} = \gamma_1 + \gamma_2$ in Eq. (11), we find $Z^M(...)$-function as follows:

$$Z^M(\nu,\Delta_{in},\gamma_1,\gamma_2,\Gamma_{tot}=\gamma_1+\gamma_2) = \frac{\gamma_1}{\Gamma_{tot}} \frac{1}{\{1+\frac{\nu^2}{\Gamma_{tot}^2\Delta_{in}^2}[\frac{1}{4}(\Gamma_{tot}-2\Pi(\nu)\Delta_{in})^2 + \Pi^2(\nu)\nu^2]\}}, \qquad (19)$$

We take into account also large enough spectral detunings in comparison with spectral width of the signal field $|\Delta_m|>>\delta\omega_s$ so that $\Pi(\nu) \cong \Pi(0) + \nu\Pi'(0) + \nu^2\Pi''(0)/2 + ...$ (where $\Pi(0) = 1 + \sum_{m=1}^{M}\Omega_m^2/\Delta_m^2$, $\Pi'(0) = \sum_{m=1}^{M}\Omega_m^2/\Delta_m^3$). By taking into account the first *matching condition* $\Gamma_{tot} \cong \gamma_1$ in (19), the QM efficiency will be close to 100% within broader spectral range if *second spectral matching condition* is held

$$\Delta_{in} = \Delta_{opt} = \frac{\Gamma_{tot}}{2\Pi(0)} \cong \frac{\gamma_1}{2[1+M\Omega^2/\Delta^2]}, \qquad (20)$$

where we have assumed for simplicity $N_m = N$, $g_m = g$, $\Delta_1 = -\Delta_2,...,\Delta_{M-1} = -\Delta_M$, $|\Delta_m|=\Delta$, $\Omega_m^2 = \Omega^2$.

In accordance with Eq. (19) and condition (20), $Z^M(...)$-function shows almost flat spectral behavior of the quantum efficiency close to unity in the frequency domain near $\nu=0$ that is determined by a slow quadratic ($\sim \nu^4$) decreasing of $Z^M(...)$-function from the unity. Eq. (20) demonstrates that increasing of the processing nodes number M reduces an operative spectral range $\Delta_{in}$ of QM for storage of the external light fields as it is presented in Fig.1 for number of processing nodes M=0,2,4,8,16. As seen from Fig.1, we can arise $\Delta_{in}$ by increasing the spectral detuning $\Delta$ even for large number M. Since $\Delta$ is limited by a free space interval of resonator, there is a maximum allowable number of processing nodes. Thus, by adjusting appropriate spectral parameters and Rabi frequencies of the processing nodes we can reach almost ideal quantum storage within the spectral range $\Delta_{in}$ as it is depicted in Fig.2.

We stress a principle advantage of the proposed here multi-mode QED cavity QM for quantum RAM of quantum computer where operation error should be smaller than $10^{-4}$ [27] with respect to the QMs based on the photon echo QMs for travelling light modes [21,29,31-33] since here 100 % efficiency occurs only for infinite optical depth of the coherent resonant atomic system (αL>>1). Thus, the highly efficient multi-mode integrated QM opens a door for practical application in quantum storage and processing with many photon qubits.

### III. QUANTUM TRANSFER BETWEEN QM AND PROCESSING NODES

Let's consider a principle scheme of the QC with four nodes. The circuit contains the multi qubit QM) (0-nd node) and three processing nodes. QM node is loaded in gradient magnetic field providing an inhomogeneous broadening of atomic frequencies $\Delta_{in} >> \delta\omega_s$ with central atomic frequency coinciding with the circuit frequency $\omega_{QM} = \omega_0$. Three processing nodes have N atoms in each node with equal frequencies $\omega_{1,2,3}$ within each node which are tuned far away from the frequency $\omega_0$. Below we demonstrate a perfect transfer of arbitrary qubit between QM node and one of the processing nodes (1-st node) in the closed QED circuit. We take into account the multi-qubit initial state encoded in the *n* temporally separated photon modes when each k-th temporal mode is stored in the QM node at the moment of time t≅$t_k$ (where $t_1$ <$t_2$ <…<$t_k$ <$t_{k+1}$ …<$t_n$ ) as it is presented in the previous section. When the storage procedure is completed (t=τ), we switch off the inhomogeneous broadening in the QM node and we tune away the atomic frequency from resonance with the circuit frequency $\omega_{QM} \neq \omega_0$. In order to transfer one arbitrary k-th qubit state from the QM node to 1-st node we switch off the coupling with the external waveguide (that leads to $\gamma_1 \approx 0$) and then we launch rephasing of the atomic coherences in QM node by reversion of the atomic detunings for time $t \geq \tau$: $\Delta_{j,1}(t \geq \tau) \rightarrow -\Delta_{j,1}$. It is obvious that the k-th qubit state will be rephased at $\tilde{t}_k = t_k + 2(\tau - t_k) = 2\tau - t_k$ - time moment of k-th echo emission. Before rephasing of k-th mode we equalize the frequencies of QM-node and of the 1-st node with the cavity frequency $\omega_{QM} = \omega_1 = \omega_0$ at the moment of time $t > 2\tau - t_{k+1} + \delta t$ (where temporal duration of each temporal mode δt<<$t_{k+1}$–$t_k$). The quantum dynamics of the k-th atomic coherence in the QM node (m=0), of atomic coherences in three processing nodes and of the circuit mode will be determined by the following system of equations

$$\frac{d}{dt} S^j_{-;m} = -(g^j_m)^* a - i\Delta_{j,m} S^j_{-;m}, \qquad (21)$$

$$\frac{d}{dt} a = \sum_{m=0}^{3} \sum_{j=1}^{N_m} g^j_m S^j_{-;m}, \qquad (22)$$

where we have ignored a weak atomic decoherence for the used timescale ($\gamma_{21} t << 1$).

In accordance with the previous section, we can choose the parameters of 2-nd and 3-rd nodes which provide a resonant interaction of QM node and 1-st processing node with the cavity field mode. Here, we take the frequencies of 2-nd and 3-rd processing nodes tuned asymmetrically far from the cavity frequency $\omega_3 - \omega_0 = -(\omega_2 - \omega_0) = -\Delta_2$ (assuming also an equal number of atoms and photon-atoms coupling constants in the nodes: $N_2$=$N_3$, $g_2$=$g_3$). Finally, we take into account large enough spectral detuning $\Delta_2$ where we can ignore a weak dispersion effects caused by 2-nd and 3-rd node to the interaction between QM node, 1-st processing node and cavity mode. By taking a formal solution of Eq. (21) in Eq. (22) and taking into account equal atomic frequencies and a resonance of the processing node with QM node and cavity field mode, $\Delta_{j,1} = \Delta_1 = 0$, we get the field equation

$$\frac{d}{dt} a(t) = N_0 \bar{g}_0 S^{(in)}_{-,0}(t) + N_1 \bar{g}_1 S^{(in)}_{-,1}(t) - \Omega_0^2 \int_\tau^t dt' \exp\{-\Delta_{in}(t-t')\} a(t') - \Omega_1^2 \int_\tau^t dt' \exp a(t'), \qquad (23)$$

where the initial atomic coherences of QM node $S_{-,0}(t)$ and of 1-st processing node $S_{-,1}(t)$ are

$$N_0 \bar{g}_0 S^{(in)}_{-,0}(t) = \sum_{j=1}^{N_1} g^j_0 S^{(j_0)}_{-,0}(\tau) \exp\{i\Delta_{j_0}(t-\tau)\}, \qquad (24)$$

$$N_1 \bar{g}_1 S^{(in)}_{-,1}(t) = \sum_{j_1=1}^{N_1} g^j_1 S^{(j_1)}_{-,1}(\tau). \qquad (25)$$

Below we evaluate the expectation values of the field and atomic coherences $_a\langle 0|..|\Psi_k(\tau)\rangle_a$ for initial state of atoms $|\Psi_k(\tau)\rangle_a$ in QM node that corresponds to the k-th stored temporal mode. By taking into account the initial ground state of atoms in 1-st node, we get $\langle S^{(j_1)}_{-,1}(\tau)\rangle = 0$ and $\langle S^{(in)}_{-,1}(t)\rangle = 0$ in Eqs. (23),(25). The initial atomic coherence in the QM node will be determined by the rephasing process of the k-th stored field mode state

$$\langle S^{(in)}_{-,0}(t) \rangle = \langle S^{(k)}_{-,0}(t - \tilde{t}_k) \rangle, \qquad (26)$$

$$\bar{g}_0 \langle S_{-,0}^{(k)}(t-\tilde{t}_k) \rangle = N_0^{-1} \sum_{j_0=1}^{N_0} g_0^j \langle S_{-,0}^{(j_0)}(t_k) \rangle \exp\{i\Delta_{j_0}(t-\tilde{t}_k)\}, \qquad (27)$$

where we have fixed rephasing of the k-th atomic coherence $\langle S_{-,k}^{(1)}(t_k) \rangle$ in QM node.

By using a Fourier transformations for the field mode $a(t) = \int_{-\infty}^{\infty} a(\nu)e^{-i\nu t} d\nu$ and atomic coherences, we get a general solution of the field Eq. (23) as follows

$$<a(t)> = N_0 \bar{g}_0 \int_\tau^t dt'' \Gamma(t-t'') < S_{-,0}^{(k)}(t''-\tilde{t}_k) >, \qquad (28)$$

$$\Gamma(\tau) = i \int_{-\infty}^{\infty} \frac{d\nu}{2\pi} \frac{(\nu+i\Delta_{in})\nu e^{-i\nu\tau}}{\prod_{m=1}^{3}(\nu-\nu_m)}, \qquad (29)$$

where $\nu_1 = -i(\Delta_{in}-2n)$, $\nu_2 = -(S+in)$, $\nu_3 = S-in$, $n = \Delta_{in}/3 - b/3$, $p = \Omega_o^2 + \Omega_1^2 - \Delta_{in}^2/3$, $D = (p/3)^3 + (q/2)^2$, $q = \Delta_{in}[\Omega_1^2 + 2\Delta_{in}^2/27 - (\Omega_o^2+\Omega_1^2)/3]$, we have here for $p > 0$: $b = 3(\upsilon-u)/2$, $S = \sqrt{3}(u+\upsilon)/2$, $u = \sqrt[3]{\sqrt{D}-q/2}$, $\upsilon = \sqrt[3]{\sqrt{D}+q/2}$ and for $p < 0$: $b = 3R \cdot Ch(\varphi/3)$, $S = \sqrt{3}R \cdot Sh(\varphi/3)$, $R = \sqrt{-p/3}$, $Ch(\varphi) = q/2R^3$.

All three roots of the cubic equation related to Eq. (23) $\nu_{1,2,3}$ have negative imaginary parts revealing a causality in the field Eq. (23) that leads to $\Gamma(\tau<0)=0$. By taking into account (28) and temporal properties of response $\Gamma(\tau)$-function, we introduce a k-th atomic coherence with a temporal mode related to the response $\Gamma(\tau)$-function in the following way

$$< S_{-,0}^{(k)}(t''-\tilde{t}_k) >_\Phi = \xi \int dt' \Gamma(\tilde{t}_k-t'') = \xi \Phi(\tilde{t}_k-t''), \qquad (30)$$

$$\Phi(\tau) = \frac{F}{S(b^2+S^2)} \{\sin\{\alpha-S\tau\}e^{-n\tau} - \sin\{\alpha\}e^{-(\Delta_{in}-2n)\tau}\}, \qquad (31)$$

where the mode $\Phi(\tau<0)=0$ and we call this mode as *a temporal self-mode* of QC depicted in Fig.3, $F = \sqrt{(S^2+b(\Delta_{in}-n))^2 + S^2(\Delta_{in}-n-b)^2}$, $tg\{\alpha\} = \frac{S(\Delta_{in}-n-b)}{S^2+b(\Delta_{in}-n)}$ and $\xi$ is a normalizing constant. We find the constant $\xi$ by assuming that each k-th temporal mode contains a single photon wave packet. Here, the field mode is given by the equation

$$\xi\Phi(\tau) = \int_{-\infty}^{\infty} \frac{d\nu}{\pi} \frac{\Delta_{in}}{\nu^2+\Delta_{in}^2} f(\nu)e^{-i\nu\tau}. \qquad (32)$$

By using Eqs. (29), (30) and (32), we find

$$f(\nu) = \xi \frac{(\nu+i\Delta_{in})(\nu^2+\Delta_{in}^2)}{2\Delta_{in} \prod_{m=1}^{3}(\nu-\nu_m)}, \qquad (33)$$

where $|f(\nu_j)|^2$ determines a probability to find j-th atom of the QM node in the excited state, i.e.

$$\sum_{j=1}^{N_o} |f(\nu_j)|^2 = N_o \int_{-\infty}^{\infty} \frac{d\nu}{\pi} \frac{\Delta_{in}}{\nu^2+\Delta_{in}^2} |f(\nu)|^2 = 1. \qquad (34)$$

That leads to the following normalizing constant $\xi = \sqrt{2\Delta_{in}/(N_0 K)}$, where

$$K = -i\sum_{n=1}^{3} \frac{(\nu_n^2 + \Delta_{in}^2)^2}{\prod_{m \neq n}(\nu_n - \nu_m)\prod_{m'=1}^{3}(\nu_n - \nu_{m'}^*)}. \qquad (35)$$

By using Eqs. (30),(31), we find a complete depopulation of the k=th field mode at $t \cong \tilde{t}_k$:

$$<a(t=\tilde{t}_k)>_\Phi = \xi N_0 \bar{g}_0 \{\Phi^2(0) - \Phi^2(\tilde{t}_k - \tau)\} = \xi_b N_0 \bar{g}_0 \Phi^2(0) = 0, \qquad (36)$$

where we have assumed a large dephasing of the initial k-th atomic coherence in QM node $<S_{-,0}^{(k)}(\tau - \tilde{t}_k)>_\Phi \cong 0$. We note that the excited field mode $<a(t)>_\Phi$ is a continuous function of time in contrast to $\Gamma(t-t')$-function at $t-t'$.

In order to realize a perfect quantum transfer of k-th temporal field mode, we have to find optimal parameters of the atomic nodes providing an ideal mode transfer to the processing node at some fixed moment of time. By using Eqs. (28), (30), we obtain the atomic coherence of 1-st processing node as follows

$$<S_1(t)>_\Phi = \xi \bar{g}_1 N_0 \bar{g}_0 \int_{-\infty}^{t} dt' \int_{-\infty}^{t'} dt'' \frac{\partial}{\partial t''} \Phi(t'-t'')\Phi(\tilde{t}_k - t'')\Big|_{t=\tilde{t}_k} = -\xi \frac{\bar{g}_1}{\bar{g}_0}\Omega_0^2 P_1(\tilde{t}_k), \qquad (37)$$

where $P_1(\tilde{t}_k) = \int_0^\infty d\tau \Phi(\tau)^2$.

Quantum efficiency of the k-th mode transfer at $t = \tilde{t}_k$ to the 1-st processing node will be given by

$$Q_1(\tilde{t}_k) = N_1 |<S_1(\tilde{t}_k)>_\Phi|^2 = \frac{2\Delta_{in}}{K}\Omega_0^2 \Omega_1^2 P_1^2(\tilde{t}_k). \qquad (38)$$

Numerical evaluation of the quantum efficiency $Q_1(\tilde{t}_k)$ for various atomic parameters of QM and processing nodes ($\Delta_{in}$, $\Omega_0$ and $\Omega_1$) is presented in Fig. 3 for $\Omega_0 = 1$, $\Omega_1 = 0.3$. Here, we have made the calculations in the units of $\Omega_0$-Rabi frequency and inhomogeneous broadening in spectral region $0.2 < \Delta_{in} < 10$. As seen in Fig.3, the *temporal self-mode* has a smooth temporal shape only for intermediate spectral range $0.5 < \Delta_{in} \leq 5.5$ that can be explained in the following way. For small inhomogeneous broadening $\Delta_{in} < 0.5$, we observe an obvious long-term quantum oscillations between QM and processing node at the state transfer. However, if we increase $\Delta_{in} > 4$ at the same magnitude of $\Omega_0 = 1$, we will reduce the resonant coupling strength between the QM node and QED cavity that will increase the transfer time to the processing node. Thus, there is an optimal atomic parameters for fast and robust transfer between the QM and processing node. Also, we note an unobvious property of the state transfer revealing in a continuous increasing of the quantum efficiency $Q_1(\tilde{t}_k)$ with inhomogeneous broadening $\Delta_{in}$ (Fig.4). We must obtain high enough quantum efficiency $Q_1(\tilde{t}_k) > 0.9999$ for realization of fault tolerant quantum computer (see [20]). As it is seen in Fig. 5, we can reach the necessary quantum efficiency for $\Delta_{in} > 5.5$ where the temporal field mode has a still smooth shape and short temporal duration that seems useful for practice.

Thus, the quantum efficiency $Q_1$ can be very close to 100% for relatively large inhomogeneous broadening $\Delta_{in}$ in comparison with the quantum Rabi frequencies of the QM and processing nodes ($\Delta_{in} > \Omega_0$ and $\Delta_{in} > \Omega_1$) as it was shortly noted in [15] for quantum computer on multi-atomic ensembles. The discussed properties of the optimal *temporal self-modes* reveal experimentally achievable coupling strength of QM and processing nodes in the

QC scheme that somehow reminds the properties of optimal quantum storage processes studied recently in [34,35]. It is worth noting that the proposed integration of the QM integration can be also applied for the QC with superconducting or quantum dot processing nodes characterized only by its sufficiently high Rabi frequencies $\Omega_1$. In order to fix the transferred state in the processing node, we have to switch off abruptly at $t = \tilde{t}_k$ the resonant coupling of the cavity mode with QM and processing nodes providing a freezing of the transferred photon qubit in 1-st processing node.

In order to realize a backward transfer of the photon qubit to the QM node, we can use an interesting symmetry of Eqs. (21), (22) under the following *symmetry transformation*: $t \to -t'$, $\Delta_{j,m} \to -\Delta_{j,m}$ (i.e. inversion of time and of atomic frequency detunings) plus a sign change of the field mode operator $a \to -b$. This symmetry transformation has been proposed for general studies of the time reversibility of the photon echo QM based on the traveling field modes [19], also used for finding general quantum states of the light fields retrieval [20] and has been generalized recently for new reversible schemes in [33]. The symmetry transformation leads now to the system of equations for new operators $b(t'), S_{j_m}(t')$ which coincides completely with Eqs. (21), (22). This fact automatically leads to the following important obvious conclusion. The dynamics started at some moment of time from the processing node with empty cavity mode and rephased state of the QM node will lead to generation and storage of the temporally reversed self-temporal mode in the QM node accompanied by the depopulation of the processing node and cavity node, since this process will be a temporally reversed copy of the studied above quantum transfer.

It is worth noting that we don't need changing a sign of the field mode in the case of the completely depopulated field mode at $t = \tilde{t}_k$ where we get a temporally reversed transfer automatically in the further quantum evolution leading to perfect return of the quantum state from the processing node to the QM node. However, if we have to reverse the temporal evolution from the intermediate quantum state characterized by nonzero field mode, we can reverse the dynamics by accompanying *symmetry transformation* with an additional relative π-phase shift between the field mode and atomic coherences. The phase shift is realized by applying 2π-control light pulse at adjoin atomic transition (in particular for three level atoms with Ξ- or Λ- atomic configurations), as it have been discussed similarly for the photon echo QM in [37] and early in [38] for realization of temporal reversibility for single atomic revivals in QED cavity.

We can use the described protocols for initialization of many self-temporal modes in the QM node from the excited processing nodes in the case of a large enough temporal scale of the atomic decoherence time $\gamma_{21}^{-1}$. By using the described manipulations of the resonant frequencies of the QM and processing nodes as well as by exploiting the dephasing/rephasing processes of the atomic coherence in the QM node, one can also download the photon qubits in many processing nodes in order to realize full table of protocols for quantum processing on the photonic qubits. In particular we can perform a basic single and two qubit gates by using a physical encoding of single logical qubit on two processing nodes with controlled resonant coupling of the processing nodes through the interaction with main and additional field modes of QED cavities discussed in [15].

## IV. CONCLUSION

In this work, we have elaborated in details the multi-mode QM integration into the quantum computer scheme. Here we should be able to control the carrier frequencies of the QM and processing nodes in order to provide resonant interaction between the nodes with the cavity mode and external input light field. We have obtained modified optimal conditions for integration of efficient multi-qubit quantum memory in the quantum computer and we have revealed a perfect temporal shape of the self temporal quantum computer modes providing an almost ideal reversible transfer of the photon qubits between the quantum memory node and the arbitrary processing nodes. It should be stressed that we have theoretically demonstrated QC with QM in fon Neiman architecture using QM on inhomogeneously broadened line where we have no macroscopic dipole moment during the storage stage in contrast to QM on homogeneously broadened line. Moreover, the dephasing and rephasing of atomic coherence are performed automatically in the QM node without using any additional external fields. These properties of the described QM eliminate the relevant sources of decay and decoherence making creation of practical QuRAM more feasible.

We note that the elaborated quantum memory and quantum transport can be realized with almost 100% of efficiency for the optimal moderate parameters of the atomic ensembles. The described integration schemes open promising possibilities for practical realization of quantum protocols in the QC with limited number of processing nodes but with multi-qubit QM. In particular, the outlined framework for the efficient coupling between the QM and processing node can be also applied for superconducting and quantum dots QC.

## V. ACKNOWLEDGEMENT

The authors thank the Russian Foundation for Basic Researches under the grants # 10-02-01348 and 11-07-00465.

**Captions for the Figures**

Caption for Figure 1 a), Figure 1 b) and Figure 1 c).
Spectral properties of QM efficiency $[Z^M(\nu, \Delta_{in} = \Delta_{opt})]^2$ for QC with different number of processing nodes - M=0 (long dashing), 2 (middle dashing), 4 (short dashing), 8 (dotted), 16 (solid line); where we use spectral units with $\Gamma_{tot} \cong \Gamma_{in} \cong \gamma_1 = 1$ and $\gamma_1/\Delta = 0.3$ with atomic parameters of Eq. (20): $N_m = N$, $g_m = g$, $\Delta_1 = -\Delta_2, ..., \Delta_{M-1} = -\Delta_M$, $|\Delta_m| = \Delta$, $\Omega_m^2 = \Omega^2$.
Three cases: 1 a) $\Omega/\Delta = 0.4$, 1 b) $\Omega/\Delta = 0.25$, c) $\Omega/\Delta = 0.1$. As seen from the comparison of a), b) and c), the influence of the processing nodes (M≤16) is considerably reduced for $\Omega/\Delta \leq 0.1$.

Figure 2. Spectral behavior of QM efficiency in a broad spectral range $\Delta_{in}$.

Caption for the Figure 3 a) and Figure 3 b).
Temporal shape of self mode as a function of inhomogeneous broadening width $\Delta_{in}$ presented for two spectral range: a) $0 < \Delta_{in} < 1.8$; b) $0 < \Delta_{in} < 1.8$; where we use the units of QM node Rabi frequency $\Omega_0 = N\bar{g}_0^2 = 1$, $\Omega_1 = N\bar{g}_1^2 = 0.3$.

Figure 4. Quantum efficiency of quantum transfer from QM node to processing node as a function of inhomogeneous broadening width $\Delta_{in}$ where we use the units $\Omega_0 = N\bar{g}_0^2 = 1$, $\Omega_1 = N\bar{g}_1^2 = 0.3$.

Figure 5. High quantum efficiency of quantum transfer from QM node to processing node in the optimal spectral range of inhomogeneous broadening width $\Delta_{in}$, where we use the units $\Omega_0 = N\bar{g}_0^2 = 1$, $\Omega_1 = N\bar{g}_1^2 = 0.3$.

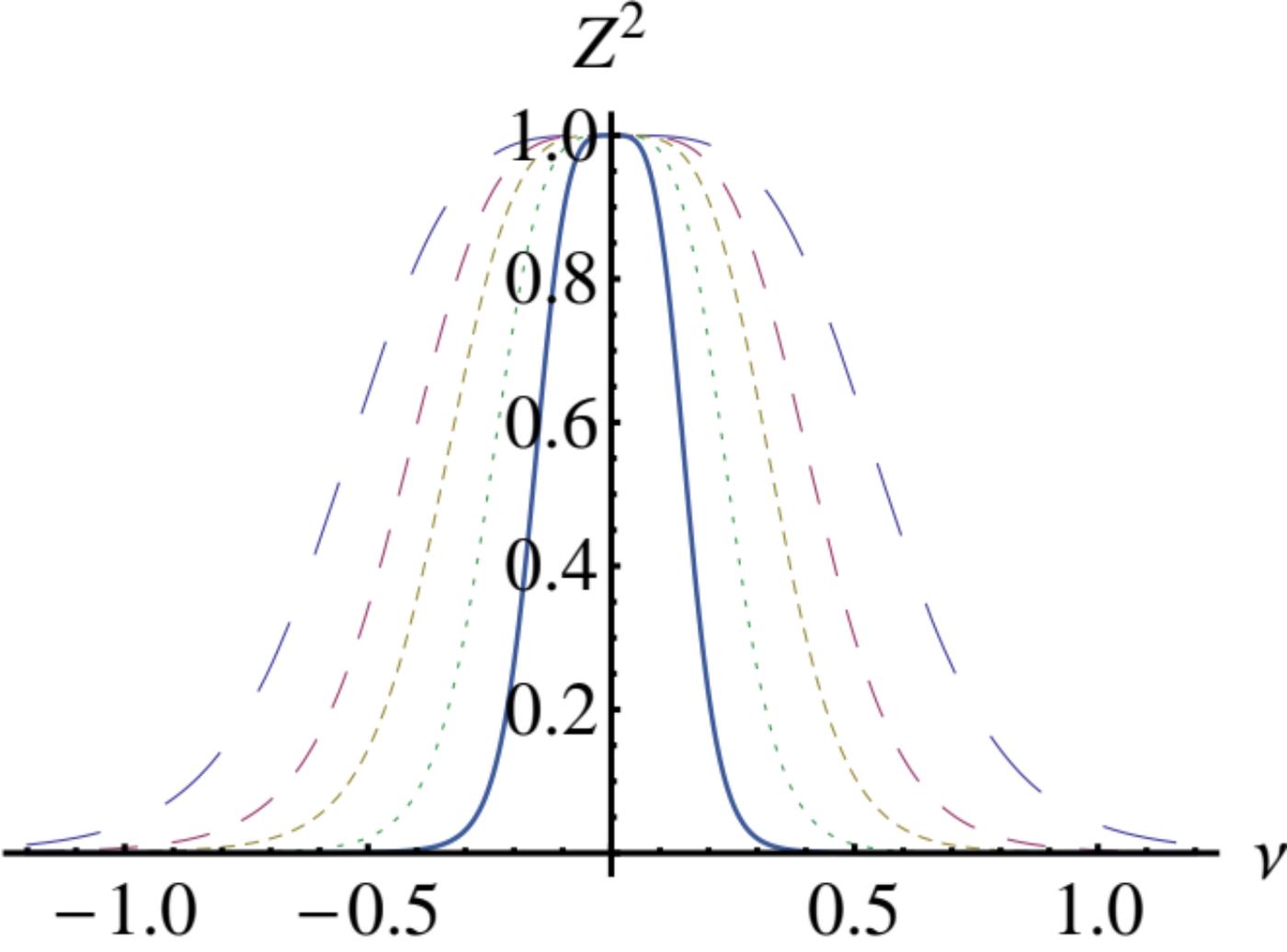

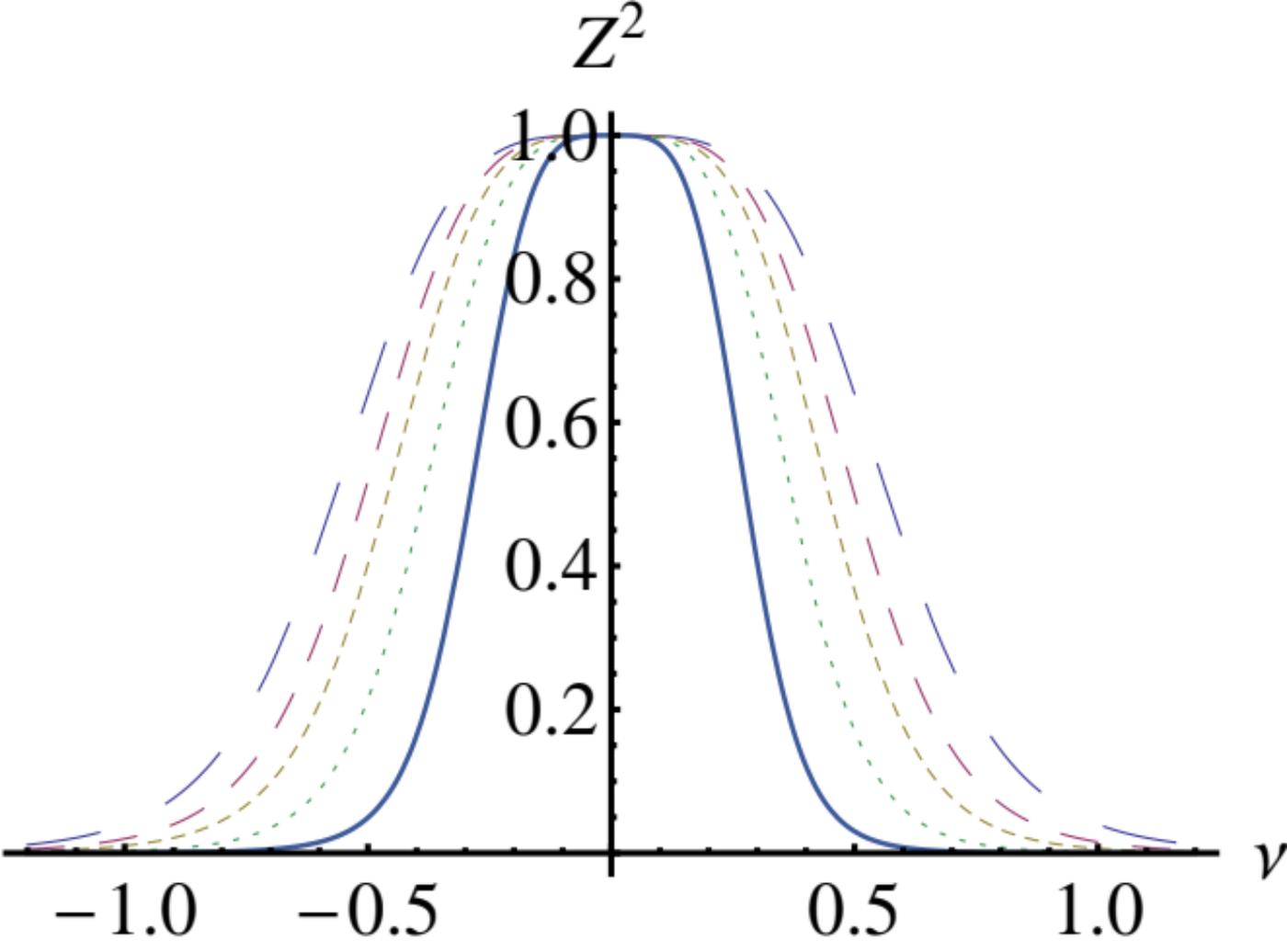

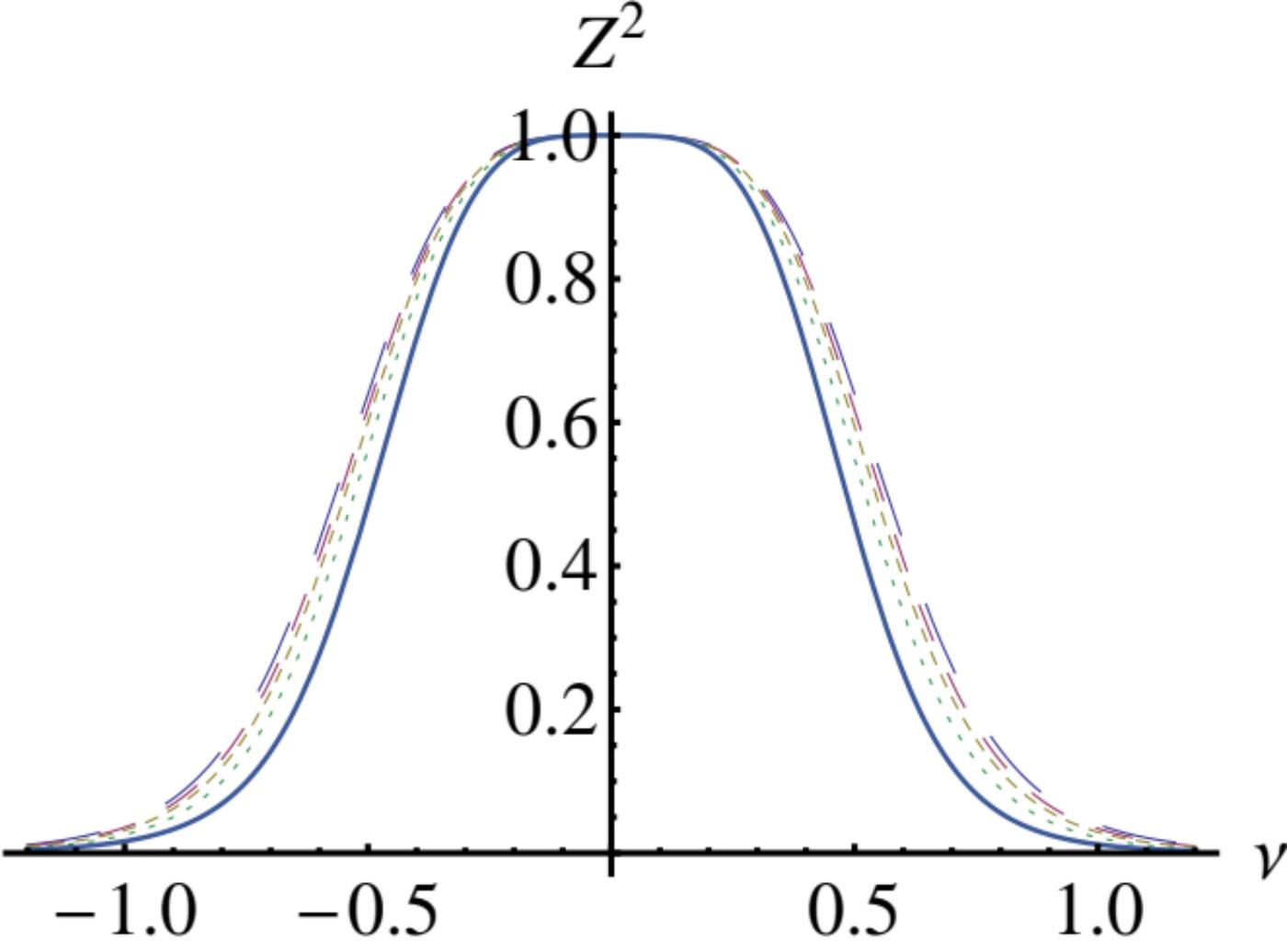

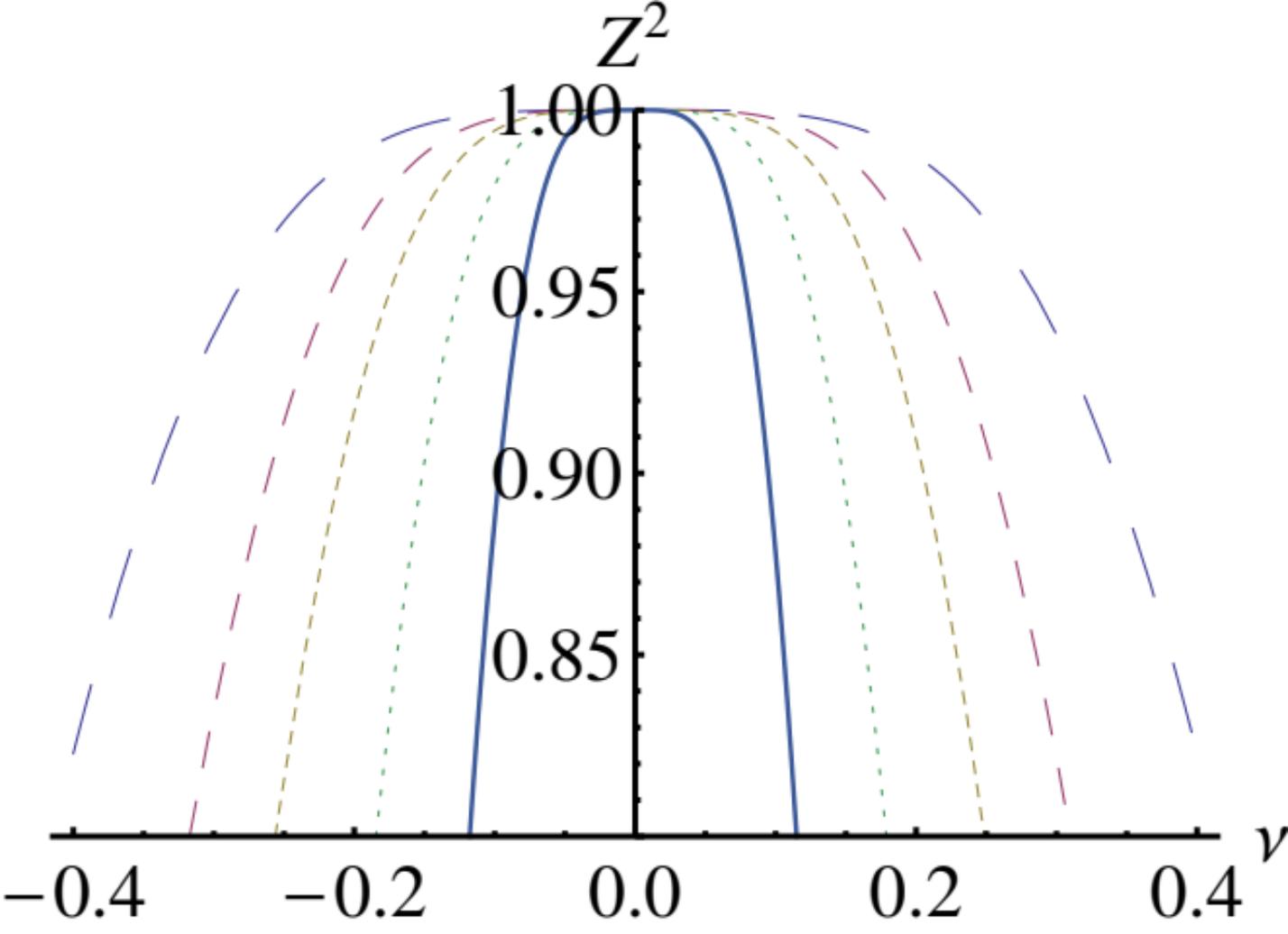

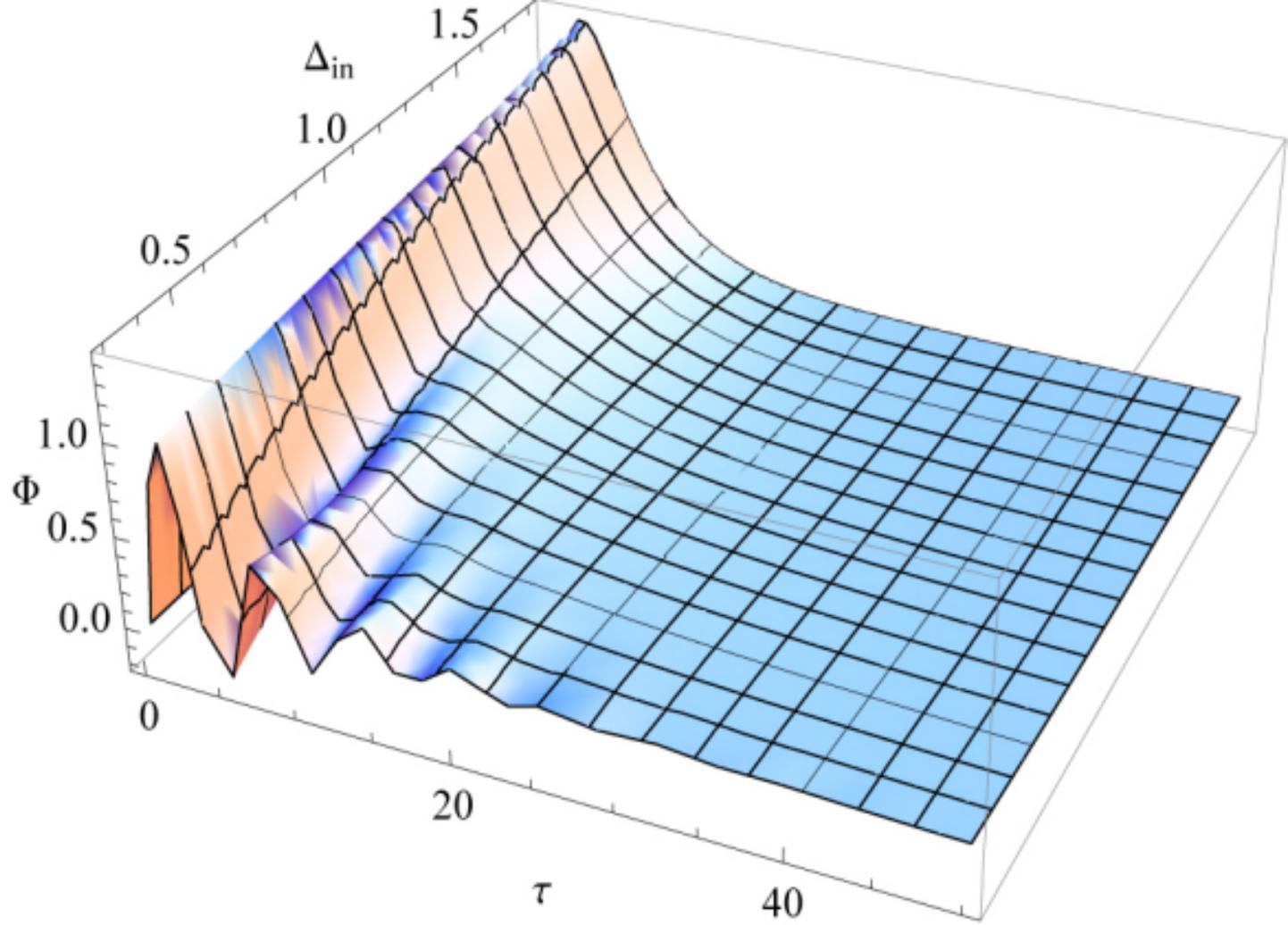

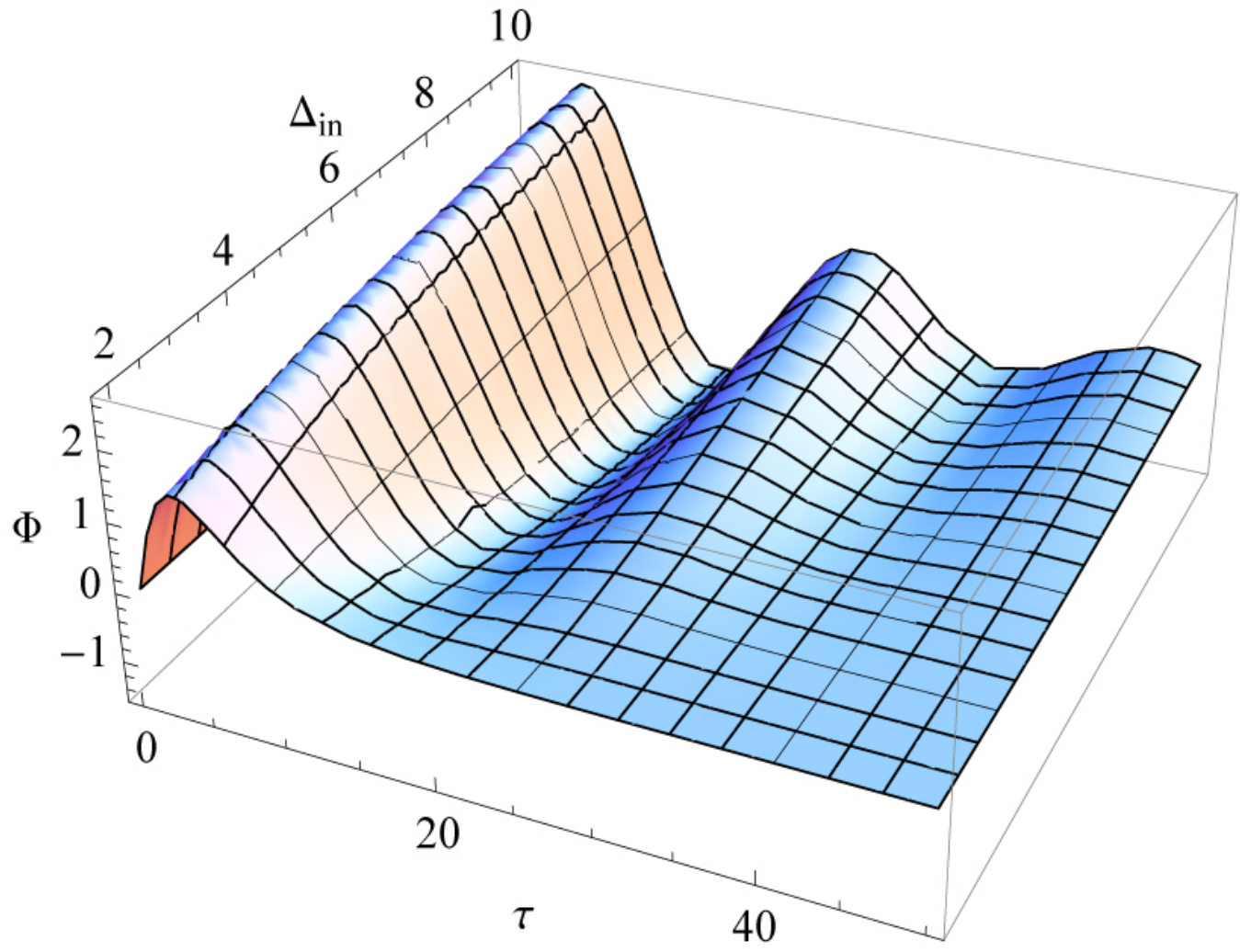

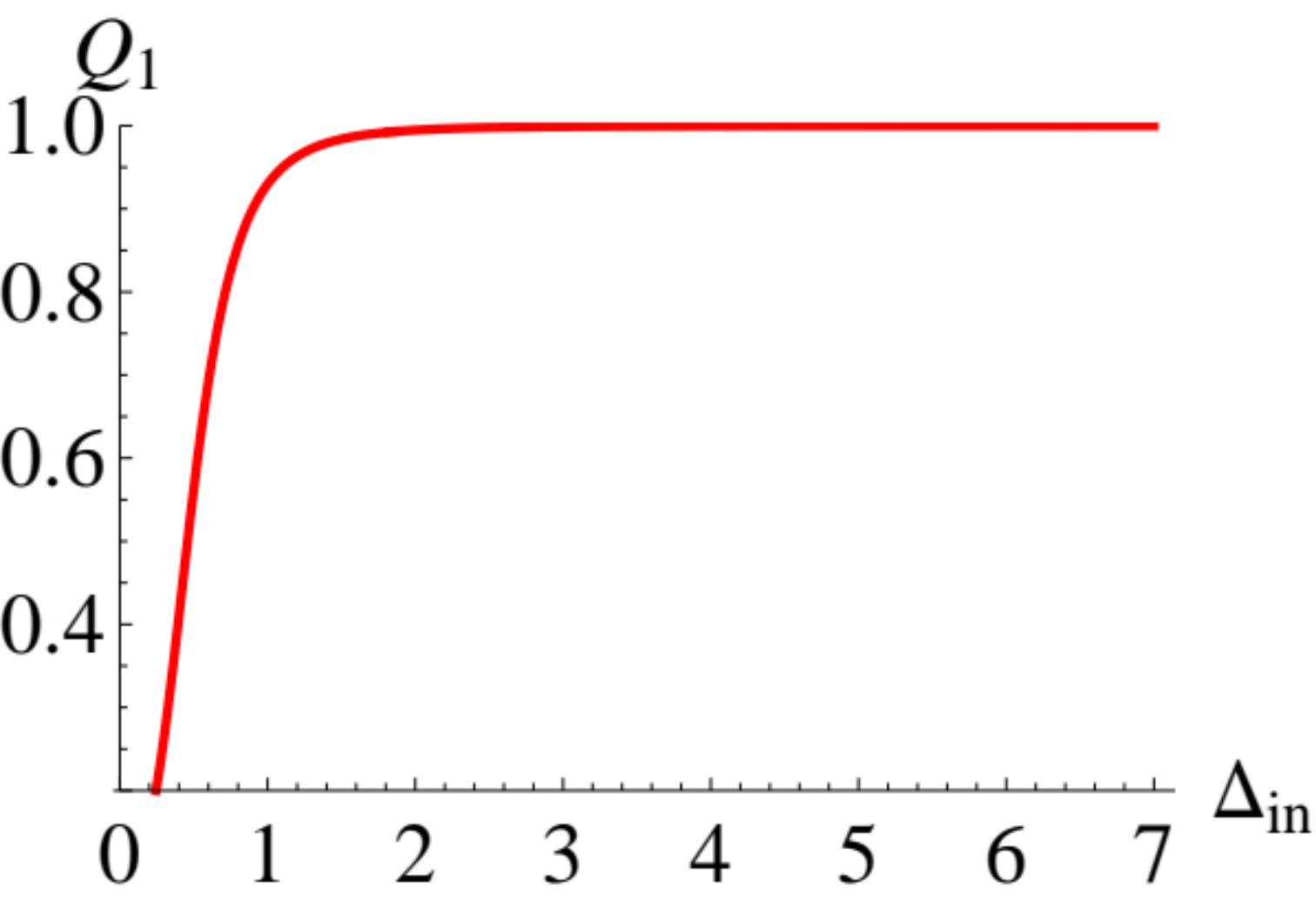

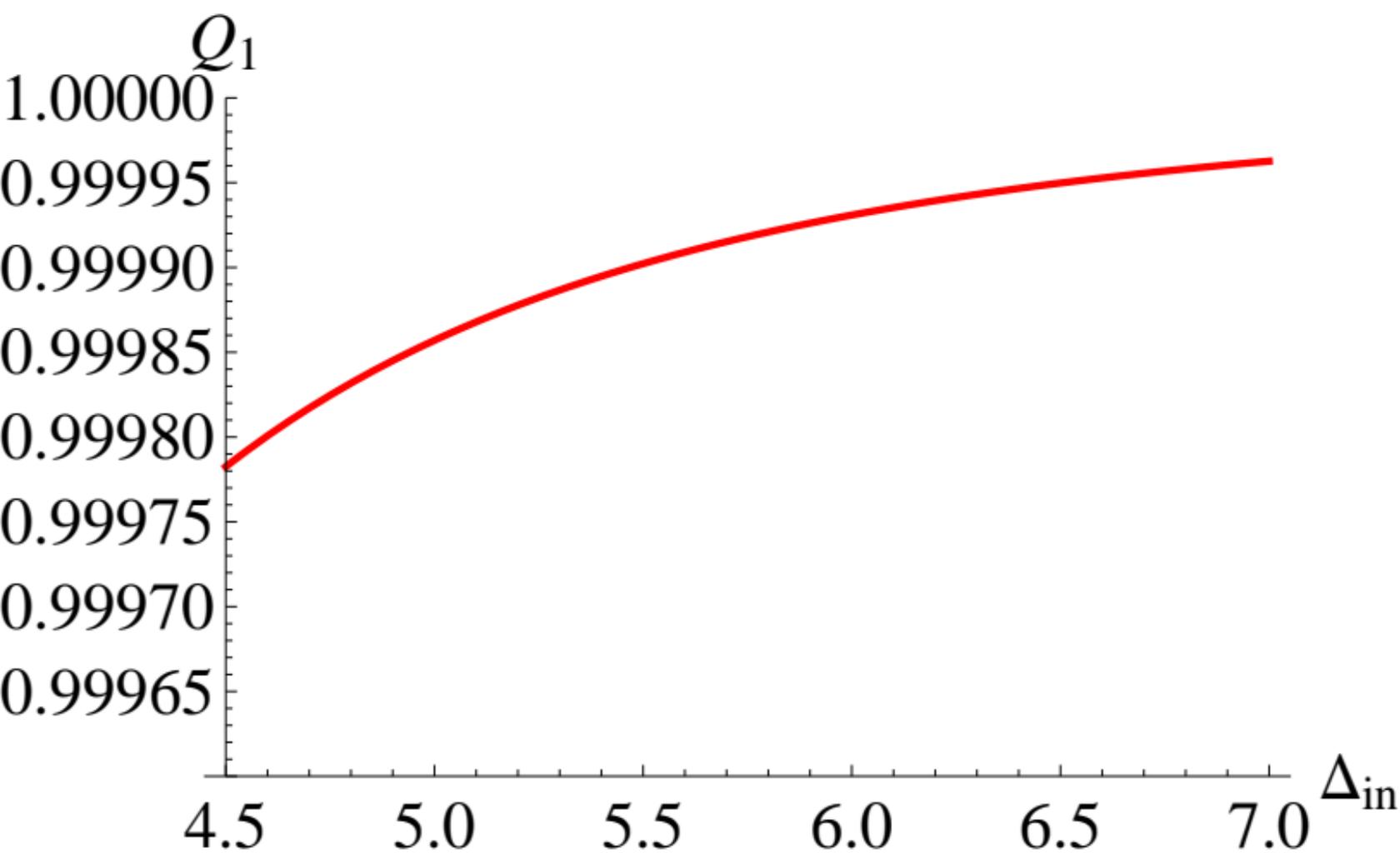